\documentclass{article}
\usepackage{spconf,amsmath,amsfonts, graphicx, amssymb,float,amssymb}
\usepackage{multirow}
\usepackage{makecell}
\usepackage{color,cite}
\usepackage{booktabs}
\usepackage{ragged2e}
\usepackage{amsmath}
\usepackage{amssymb}

\title{Speech-text based multi-modal training with \\ bidirectional attention for improved speech recognition}
\name{Yuhang Yang$^{1*}$, Haihua Xu$^{2*}$, Hao Huang$^1$, Eng Siong Chng$^3$ , Sheng Li$^4$ \thanks{$*$ Authors have equal contributions. Hao Huang is the correspondence author. }} 
\address{$^1$School of Information Science and Engineering, Xinjiang University, China \\ $^2$Bytedance AI Lab, Singapore\\
 $^3$School of Computer Science and Engineering, Nanyang Technological University, Singapore \\
 $^4$National Institute of Information and Communications Technology (NICT), Kyoto, Japan
}
\begin{document}
\ninept
\maketitle
\begin{abstract}
% The abstract should appear at the top of the left-hand column of text, about
To let the state-of-the-art end-to-end ASR model enjoy data efficiency, as well as much more % easily available 
unpaired text data by multi-modal training, one needs to address two problems: 1) the synchronicity of feature sampling rates between speech and language (aka text data); 2) the homogeneity of the learned representations from two encoders. In this paper we propose to employ a novel bidirectional attention mechanism (BiAM) to jointly learn both ASR encoder (bottom layers) and text encoder with a multi-modal learning method. The BiAM is to facilitate feature sampling rate exchange, realizing the quality of the transformed features for the one kind to be measured in another space, with diversified objective functions. As a result, the speech representations are enriched with more linguistic information, while the representations generated by the text encoder are more similar to corresponding speech ones, and therefore the shared ASR models are more amenable for unpaired text data pretraining. 
To validate the efficacy of the proposed method, we perform two
categories of experiments with or without extra unpaired text data. Experimental results on Librispeech corpus show it can achieve up to 6.15\% word error rate reduction (WERR) with only paired data learning, while 9.23\% WERR when more unpaired text data is employed\footnote{Source code: https://github.com/yuhangear/Multi-modal-learning.git}.

\end{abstract}

\section{Introduction}\label{sec:intro}
End-to-end (E2E) automatic speech recognition (ASR) framework~\cite{chan2016listen,2017State_icassp,zhang2020transformer,INTERSPEECH-conformer-2020,zhang2022non} has now come into predominance in both research and product areas \cite{he2019streaming,sainath2020streaming,li2022language} thanks to its efficacy in modeling capacity, as well as compactness. However, one of the limitations of E2E ASR modeling is its insatiable data-hungry~\cite{li2022massively}. To train a decent ASR system, the rule of thumb is always the more data the better.

To get more data, one would first consider collecting more human-transcribed data, the so-called paired data. Unfortunately, such data comes with high costs. As a result, ASR models are usually trained with limited paired data. The alternative is to get more unpaired data at a lower cost instead, in terms of either unpaired speech data or unpaired text data accordingly.
For unpaired speech data exploitation, one can employ unsupervised pretraining~\cite{ling2020deep,liu2020mockingjay,hsu2021hubert,zheng2021wav} or self-training~\cite{chung2019unsupervised,baevski2019vq,baevski2020wav2vec,chen2022wavlm,baevski2022data2vec} to yield improved ASR performance, while to take advantage of unpaired text data, people have many options for obtaining better ASR models.

In order to well exploit text data, one of the simplest ways is to employ text data to train an external language model (LM)~\cite{chorowski2016towards_lm,sriram2017cold_lm} that is fused with ASR system, yielding improved results. Besides, given a unpaired text data set, people can employ a text-to-speech (TTS) system to generate synthesized paired speech-text data~\cite{hori2019cycle,wang2020improving_tts,text-tts-2020-ibm}. However, the challenge is to obtain an off-the-shelf TTS system yielding diversified speech data is a nontrivial task.
%\textcolor{red}{What have they done and what conclusion has been reached?}

More recently, multi-modal training has been widely explored to realize training an ASR model with both speech and text (either paired or unpaired) data simultaneously~\cite{fb2021general,2021Optimizing,google2022MAESTRO}. The difficulties here lie in two aspects: 1) The synchronicity of feature sampling rates between speech and text/language, namely, speech sampling rate is much faster than language ones, and hence how to synchronize them is a problem, denoted as \textbf{AliProblem-1} for brevity; 2) The homogeneity of the learned representations from two encoders, that is, since the ASR encoder hidden representations have different distributions with those obtained from the text encoder, how to make the two representations similar is also a problem, and it is denoted as \textbf{AliProblom-2}. 

For the above-mentioned multi-modal training, \cite{2021Optimizing} employs a conventional HMM-DNN model to obtain phone level alignment for the transcript of the paired data, and the duration estimation model is used for the unpaired text data to solve \textbf{AliProblem-1}. To address the \textbf{AliProblom-2}, one can introduce diverse objective loss functions, such as masked LM (MLM), connectionist temporal classification (CTC), as well as cosine distance loss functions, etc., to make the two learned representations closer to each other.

\begin{figure}[t]
 \centering
 \includegraphics[width=0.9\linewidth]%{multi-modal-biAM-arch-1024-02.drawio}
 {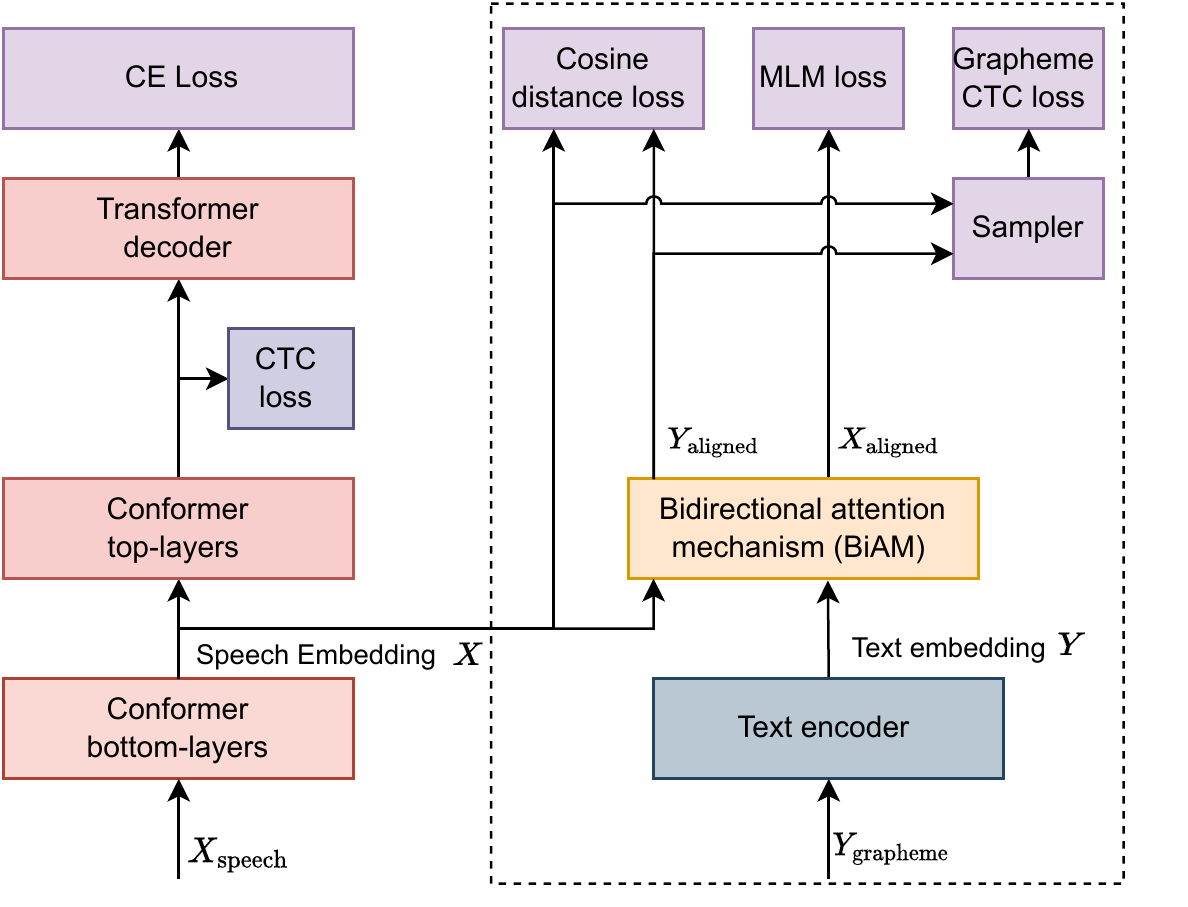}
 \caption{Speech-text based multi-modal learning framework with Bidirectional attention mechanism (BiAM). After training, all the stuff in the dashed-line box will be removed.} 
 \label{fig:multi-modal-framework} 
\end{figure}

In this paper, we propose a novel speech-text based multi-modal training approach to boost ASR performance, using a modified bidirectional attention mechanism~(BiAM)~\cite{li2022neufa} that facilitates the solution of both \textbf{AliProblom-1} and \textbf{AliProblom-2} with a joint training manner. The framework of the proposed method is illustrated in Figure~\ref{fig:multi-modal-framework}.
By BiAM, we can mutually transform one kind of representation (aka embedding) into another representational space.
Specifically, we can transform language representation (aka text embedding) into speech space, as well as transform speech representation (aka speech embedding) into language representational space. By such a transformation, we can solve the \textbf{AliProblom-1}. Meanwhile, we employ a series of loss functions, such as CTC loss, cosine distance losses, as well as MLM loss, to make the two transformed features
closer, % closer. Consequently, 
hence addressing the \textbf{AliProblom-2}. 
Concretely, once we employ the BiAM to transform the text embedding into speech space, cosine distance loss is employed to address the two feature similarity issues, the text encoder is learned to generate embeddings more appropriate for the speech encoder. Conversely, the transformed speech embedding into language representational space is measured with CTC and MLM losses, respectively, such that the bottom layer of the ASR encoder is learned to extract embeddings enriched with linguistic information. 

Our contributions can be summarized as follows: 1) To the best of our knowledge, we are the first to employ the bidirectional attention mechanism for speech-text-based multi-modal training to boost ASR performance. 
2) To train text encoder, we advocate grapheme instead of phoneme sequence to learn text encoder, which makes the proposed method language agnostic.
3) We demonstrate its efficacy on Librispeech data with diverse configurations.

\section{RELATION TO PRIOR WORK}\label{sec:prior}
Speech-text based multi-modal training for end-to-end ASR has become popular for a while~\cite{jhu2018-mmda,mit-2019-asru,2021Optimizing,fb2021general,deng2022cif, speecht52022,chen2022tts4pretrain, google2022MAESTRO}. \cite{fb2021general} directly merge the two embeddings generated from both encoders to train the shared ASR encoders.
To solve both problems as mentioned, \cite{mit-2019-asru} proposed to use the embeddings from text encoder as query while speech embeddings from ASR encoder as value to perform attention as a kind of speech-text alignment. \cite{deng2022cif} apply a CIF framework~\cite{dong2020cif} to generate phoneme-level embeddings from speech embedding, realizing text-speech alignment. 
\cite{google2022MAESTRO} proposed to employ RNNT-T to generate alignments between the text and speech encoder output.
\cite{2021Optimizing} proposed to 
use HMM-TDNN aligned phone sequence as the input to train a text encoder from which the output embedding is generated. Besides, \cite{2021Optimizing} also employed CTC loss to make both embeddings similar.

\section{Methodology}\label{sec:sys-frame}
\subsection{Multi-modal learning framework}\label{sub:framwork}
The whole framework is illustrated in Figure~\ref{fig:multi-modal-framework}, 
which is composed of three modalities, one is Conformer-based~\cite{INTERSPEECH-conformer-2020} 
ASR model, and the second is text encoder using Transformer, while the third is the modified bidirectional attention module~\cite{li2022neufa}, namely BiAM, which accepts both speech and text encoder embeddings as the inputs. 

The entire network is trained with two category losses, one is the ASR loss function, and the others are loss functions denoted as $\mathcal{L}_{\text{ALI}}$ facilitating 
the alignment optimization between two embeddings with BiAM. For clarity, the overall losses are expressed as follows:

 \begin{align}
 & \mathcal{L}_{\text{multi-modal}} = \mathcal{L}_{\text{ASR}} +
 \alpha \mathcal{L}_{\text{ALI}} 
 \label{equ:total-loss} \\
 & \mathcal{L}_{\text{ASR}} = \lambda \mathcal{L}_{\text{CTC}} + (1-\lambda)
 \mathcal{L}_{\text{Attention}} \label{equ:asr} \\
 & \mathcal{L}_{\text{ALI}} = \mathcal{L}_{\text{cd}} (\bf{Y}_{\text{aligned}}, {\bf X} ) + 
 \mathcal{L}_{\text{MLM}} (\bf{X}_{\text{aligned}}, {\bf Y}_{\text{grapheme}}) \nonumber \\ & ~~~~~~~~~~~~~~~~~~~~~~~~+\mathcal{L}_{\text{gCTC}} (\text{Sampler}({\bf X}, {\bf Y}_{\text{aligned}}), {\bf Y}_{\text{grapheme}})
 \label{equ:alg}
 \end{align}
where 
${\bf Y}_{\text{grapheme}} \in {\mathbb R}^{n_2}$ is the grapheme sequence generated from the input text data,
and $n_2$ is the sequence length in grapheme. Correspondingly, we denote the speech embedding length as $n_1$ in the following. Besides, we fix $\alpha=0.1$, and $\lambda=0.3$ in the following experiments.

Similarly in Equation~\ref{equ:alg}, 
both ${\bf X} \in \mathbb {R}^{n_1 \times d} $ and ${\bf Y} \in \mathbb{R}^{ n_2 \times d}$ are embedding sequences of the ASR and text
encoders respectively\footnote{For simplicity, we ensure the dimension of speech and text embeddings are equal to $d$.}, while ${\bf X}_{\text{aligned}} = \text{BiAM}(\bf{X})$, and ${\bf Y}_{\text{aligned}} = \text{BiAM}(\bf{Y})$ with ${\bf X}_{\text{aligned}} \in {\mathbb R}^{n_2 \times d}$, ${\bf Y}_{\text{aligned}} \in {\mathbb R}^{n_1 \times d}$, and again $n_1$ and $n_2$ being speech and grapheme embedding length respectively.
One can refer to Section~\ref{sub:biam} for the details of the explanation of the BiAM.

Besides, in Equation~\ref{equ:alg}, ``cd" stands for cosine distance, and gCTC stands for grapheme CTC. For gCTC training, we employ a ``Sampler" to sample both $\bf{X}_{\text{aligned}}$ and $\bf{Y}^{\prime}$ for each mini-batch training. As mentioned, the MLM in Equation~\ref{equ:alg} refers to masked LM.

We note that the speech embeddings are from the bottom 8th layer of the Conformer in practice, while the grapheme embeddings are output from the final layer of the text encoder instead.
Finally, after training, only the ASR modality serves for recognition in Figure~\ref{fig:multi-modal-framework}.

\subsection{Bidirectional attention mechanism}\label{sub:biam}
 To solve the alignment problem between the length of the paired speech and text embeddings (\textbf{AliProblom-1} here), \cite{li2022neufa} recently proposed a bidirectional attention
mechanism (BiAM)
realizing a neural forced-alignment (NeuFA) method. Inspired by~\cite{li2022neufa}, we propose a simpler one for the speech-text multi-modal training. Specifically,
we make $K_1= V_1$ and $K_2=V_2$, as well as the compatibility function being defined as matrix dot product in \cite{li2022neufa}. In other words, 
we do not generate key-value pairs, and we directly use text and speech embeddings for dot product operation to generate the shared attention matrix instead.
Consequently, the BiAM is implemented as follows. 
Rewrite speech embedding sequence ${\bf{X}} \in \mathbb {R}^{n_1\times d}$ as ${\bf{X}}^{n_1\times d}$ for notational clarity.
Likewise, the corresponding text embedding sequence ${\bf Y} \in \mathbb {R}^{n_2 \times d}$ is rewritten as ${\bf{Y}}^{n_2\times d}$, and
$n_1\!\!\neq\!\!n_2$. To begin with the bidirectional attention transformation, we first obtain the shared attention matrix $\bf{A}$ as: 
\begin{equation}
 { \bf{A}} = {\bf{X}}^{n_1 \times d} \times {\bf{Y}}^{d \times n_2} 
 \label{eq:att-mat}
\end{equation}
where ${\bf{A}} \in \mathbb {R}^{n_1\times n_2}$. Then we perform the softmax operation on $\bf A$ and 
${\bf A}^{T}$ to obtain:
\begin{equation}
 { \bf W}_{12}, {\bf W}_{21} = {\rm softmax}({\bf A},{\bf A}^{T})
 \label{eq:align-mat-after-softmat}
\end{equation}
where ${\bf W}_{12} \in {\mathbb R}^{n_1 \times n_2}$, and ${\bf W}_{21} \in {\mathbb R}^{n_2 \times n_1}$. Now, we can obtain two outputs as aligned embedding with the following transformation:
\begin{align}
 &{\bf X}^{n_2\times d}_{\text{aligned}} = {\bf W}_{21} \times {\bf X}^{n_1 \times d} 
 \label{eq:x-align} \\
 &{\bf Y}^{n_1 \times d}_{\text{aligned}} = {\bf W}_{12} \times {\bf Y}^{n_2 \times d}
 \label{eq:y-align}
\end{align}
where ${\bf X}^{n_2\times d}_{\text{aligned}}$ and ${\bf Y}^{n_1 \times d}_{\text{aligned}}$ are the two final outputs by the BiAM. From Equations \ref{eq:x-align} and \ref{eq:y-align}, BiAM realizes two transformations ${\bf W}_{12}$ and ${\bf W}_{21}$. The latter transforms the speech 
embeddings, yielding the ``aligned" speech sequence with the same length as the grapheme embedding length $n_2$. Likewise, the former do the opposite operation, with the ``aligned" text sequence having the same length as the corresponding speech $n_1$. Consequently, they are comparable with diverse loss functions, such as $\mathcal{L}_{\text cd}$, $\mathcal{L}_{\text{MLM}}$, and $\mathcal {L}_{\text gCTC}$ ,etc. in Equation~\ref{equ:alg}.

\begin{figure}[t]
 \centering
 \includegraphics[width=0.8\linewidth]{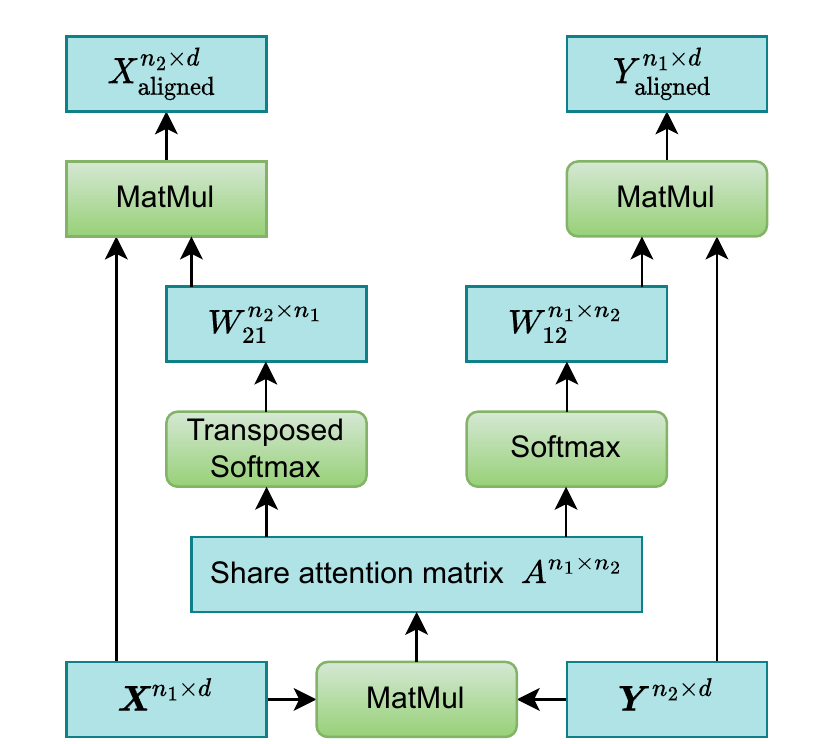}
 \caption{The diagram of bidirectional attention mechanism (BiAM), where ${\bf X}^{n_1\times d}$ and ${\bf Y}^{n_2 \times d}$ are speech and text embedding sequences respectively.} 
 \label{fig:biam}
\end{figure}
\label{ssec:subhead}

The key point of the BiAM lies in the so-called compatibility function definition in ~\cite{li2022neufa}. Here, it is defined as two embedding sequence dot product computation as Equation~\ref{eq:att-mat}, which actually is the pair-wise dot product distance between the two embedding sequences.
Once the matrices $\bf A$ and ${\bf A}^{T}$ are transformed to posterior matrix using softmax operation, they can act as attention mechanism on the input embeddings, yielding a kind of forced alignment. The details of the BiAM computation are illustrated in Figure~\ref{fig:biam}.

\subsection{Training process}\label{sub:train-process}

The whole network in Figure~\ref{fig:multi-modal-framework} is trained using 
Equation~\ref{equ:total-loss} as The loss function. In practice, we first train the network
with paired speech-text data, and both the ASR model and text encoder are jointly trained.
During this stage, the cosine distance loss $\mathcal{L}_{\text cd}$ in Equation~\ref{equ:alg} is only employed at later training steps for the sake of stable training. 

Once the training with the paired speech-text data is finished, the embeddings generated with the bottom layer of the ASR encoder have been enriched with more linguistic information that are not only speaker but also ambiance independent, such that it leads to improved ASR performance.
Optionally, after the paired speech-text data training done,
we can employ unpaired text data to continue to train the network, where the embeddings of the text encoder are taken as input to the 8th layer of the ASR encoder.
This is possible because our text encoder is also taught how to generate grapheme embeddings that are closer to the speech ones with the corresponding losses in Equation~\ref{equ:alg}.
After the unpaired text data training, we should fine-tune the network using the paired speech-text data again.
However, for the unpaired training, since we cannot perform the BiAM-based training, and the output from the text encoder has no duration information, we just randomly replicate each grapheme embedding twice so far.

\section{Experiments}
\label{sec:exp}
\subsection{Data}\label{sub:exp-data}
 All of the experiments are conducted on the LibriSpeech~\cite{ICASSP-CTC-2018} corpus.
Train data consists of 100 hours of train clean data, as well as 960 hours of full train data.
Test sets consist of 4 data sets, namely, \texttt{dev-clean}, \texttt{dev-other}, \texttt{test-clean}, and \texttt{test-other}.
Overall we conduct two kinds of experiments. One is using 100 hours of train clean data, with or without 960 hours of transcript as unpaired text data.
The other experiments are performed on the full 960 hours of train data. 

\subsection{Modeling}\label{sub:exp-modeling}
 All experiments are conducted with Espnet toolkit~\cite{watanabe2018espnet}.
The ASR model is Conformer with 12-layer encoder and 6-layer Transformer-based decoder. We use a smaller ASR model for 100-hour clean train data, while a bigger one for the 960-hour full training data. The differences lie in the middle layer, attention and word embedding dimensions, as well as multi-head attention heads, \{1024, 256, 256, 4\} for the smaller model versus \{2048, 512, 512, 8\} for the bigger model. 
The input features are 80-dimensional filter-bank, and the output is word piece models with 5000 subwords. 
The text encoder uses Transformer framework with 3- and 6-layer for 100- and 960-hour train data respectively. The differences between smaller and bigger models are the same as those of the ASR models. 
We use 0.002 learning rate for the multi-modal training on a single GPU (v100), with the 0.1 dropout. The whole network is trained with 80 epochs, and after 70 epochs the cosine distance loss is enabled with 10 epochs continuing training.
For the grapheme-based CTC training, we sample between the aligned speech and the text embeddings in each mini-batch, with each occupying 50\% samples.
For the MLM training, we randomly mask 20\% graphemes for each utterance. 

For the unpaired text pretraining, the output text embeddings are taken as input to the 8th layer of the ASR encoder. During training, During training, only ASR decoder parameters are updated, and the remaining parameters are fixed. After that, we use a 0.001 learning rate to fine-tune the ASR network with the paired speech-text data. 

For inference, the beam sizes are 20 and 60 for the 100- and 960-hour train data, respectively.

\subsection{Results}~\label{sub:exp-result}
\vspace{-8mm}
\subsubsection{Results on the 100-hour train data}
Table \ref{tab:100-hour-paired-res} presents the results of the multi-modal training using the 100-hour train data.

\begin{table}[H]
 \centering
% \caption{Statistics information for \texttt{Train}, \texttt{Valid}, $\texttt{eval}_{\texttt{man}}$ is dominated by Mandarin and ${eval}_{sge}$ is dominated by Singapore English.}
% \caption{WER[\%] for different models using 100h training data.{Grapheme CTC} is added after the Conformer bottom blocks }
\caption{WERs(\%) of the proposed BiAM-based multi-modal training with the 100-hour train data, ``cd" refers to cosine distance loss}
 \label{tab:100-hour-paired-res}
\begin{tabular}{ p{60pt} p{29pt} p{29pt} p{29pt}p{29pt} }	
 \toprule
 & \multicolumn{2}{c}{Dev WER (\%)} & \multicolumn{2}{c}{Test WER (\%)} \\
 &\hfil clean &\hfil other &\hfil clean &\hfil other \\
 \midrule
 Baseline & \hfil 6.3 &\hfil 17.4 &\hfil 6.5 &\hfil 17.3\\
 Grapheme CTC &\hfil 6.2 &\hfil 17.1 &\hfil 6.2 &\hfil 17.0\\
 \midrule
 BiAM (w/o cd) &\hfil 6.1 &\hfil 16.9 &\hfil 6.2 &\hfil 16.6 \\
 BiAM (w/ cd) &\hfil 6.0 &\hfil 16.7 &\hfil 6.1 &\hfil 16.4 \\
 \bottomrule
\end{tabular}
\end{table}
From Table~\ref{tab:100-hour-paired-res}, the proposed method gets obvious WER reductions (WERRs) on the four test sets, namely 4.76\%, 4.02\%, 6.15\% ,and 5.20\% WERR on \texttt{dev-clean}, \texttt{dev-other}, \texttt{test-clean}, and \texttt{test-other} over the baseline respectively. Furthermore, we found that cosine distance loss is very essential to get improved results, over that case where only gCTC and MLM losses are employed.
BTW, we also compare the proposed method with a multi-task learning method, namely intermediate gCTC from the 8th layer of the Conformer encoder, named ``Grapheme CTC" in Table~\ref{tab:100-hour-paired-res}.
From Table~\ref{tab:100-hour-paired-res}, though the ``Grapheme CTC" is also very effective, the proposed method has achieved consistent performance improvement. In what follows, we abbreviate the proposed BiAM with cosine distance loss as BiAM.

Table~\ref{tab:960-unpaired-res} reports the WERs of the proposed method with the 100-hour train data using 960-hour train transcript as unpaired text data.
\begin{table}[H]
 \centering
% \caption{Statistics information for \texttt{Train}, \texttt{Valid}, $\texttt{eval}_{\texttt{man}}$ is dominated by Mandarin and ${eval}_{sge}$ is dominated by Singapore English.}
% \caption{WER[\%] for different models. transcripts of 960 hours data were used as the unpaired data.}
\caption{WERs (\%) of the proposed multi-modal training using the BiAM with the 100-hour train data, puls taking 960-hour train transcript as unpaired text data }
 \label{tab:960-unpaired-res}
\begin{tabular}{ p{60pt} p{29pt} p{29pt} p{29pt}p{29pt} }	
 \toprule
 &\multicolumn{2}{c}{Dev WER (\%)} & \multicolumn{2}{c}{Test WER (\%)} \\
 & clean & other & clean & other \\
 \midrule
 Baseline & 6.3 & 17.4 & 6.5 & 17.3\\
 \midrule
 BiAM& 6.0 & 16.7 & 6.1 & 16.4 \\
 +unpaired text & 6.0 & 16.5 & 5.9 & 16.3 \\
 % ++rich text loss & 5.9 & & & \\
 \bottomrule
\end{tabular}
\end{table}
Given the unpaired text pretraining, Table~\ref{tab:960-unpaired-res} reveals the proposed method gets further WERR on the 3 test sets of the overall 4 test sets over the paired speech-text training method(see Table~\ref{tab:100-hour-paired-res}). Specifically, the WERRs are 
4.76\%, 5.17\%, 9.23\%, 5.78\% on the four test sets over the baseline model.
We notice that the unpaired text data pretraining has limited contribution to performance improvement. We think the following reason mainly accounts for this. 
During the unpaired text pretraining, we cannot get the transform ${\bf W}_{12}$ in Eq.~\ref{eq:align-mat-after-softmat}, so that the pretraining, naively employing the embedding from the text encoder, is actually a mismatched training. 
% 1) We only fine-tune the ASR decoder; hence, the contribution is limited. 2) The phone duration model is not employed, and our double replication method is somewhat rough.

Table~\ref{tab:960-hour-paired-text} reports WERs of the proposed method using 960-hour train data.
\begin{table}[H]
 \centering
% \caption{Statistics information for \texttt{Train}, \texttt{Valid}, $\texttt{eval}_{\texttt{man}}$ is dominated by Mandarin and ${eval}_{sge}$ is dominated by Singapore English.}
% \caption{WER[\%] for different models using 960h train data}
\caption{WERs (\%) of the proposed BiAM-based multi-modal training method with 960-hour paired training data.}
 \label{tab:960-hour-paired-text}
\begin{tabular}{ p{60pt} p{29pt} p{29pt} p{29pt}p{29pt} }	
 \toprule
 & \multicolumn{2}{c}{Dev WER(\%)} & \multicolumn{2}{c}{Test WER (\%)} \\
 & clean & other & clean & other \\
 \midrule
 Baseline & 2.1 & 5.2 & 2.4 & 5.3\\
 Grapheme CTC & 2.1 & 5.2 & 2.4 & 5.2\\
 \midrule
 % +Bi-D Attention & 2.1 & 5.2 & 2.3 & 5.2 \\
 % ++Cosine Loss & 2.0 & 5.0 & 2.3 & 5.0 \\
 BiAM & 2.0 & 5.0 & 2.3 & 5.0 \\
 \bottomrule
\end{tabular}
\end{table}
What is shown in Table~\ref{tab:960-hour-paired-text} again validates the efficacy of the proposed method for speech-text-based multi-modal training. It has achieved
consistent WERR over the baseline model. The WERRs are 4.76\%, 3.85\%, 4.17\% and 5.66\% on the four test sets, respectively. Besides,
compared with the intermediate CTC-based multi-task learning method, the proposed method also has a clear improvement margin.

\begin{figure}[t]
 \centering
 \includegraphics[width=0.99\linewidth]{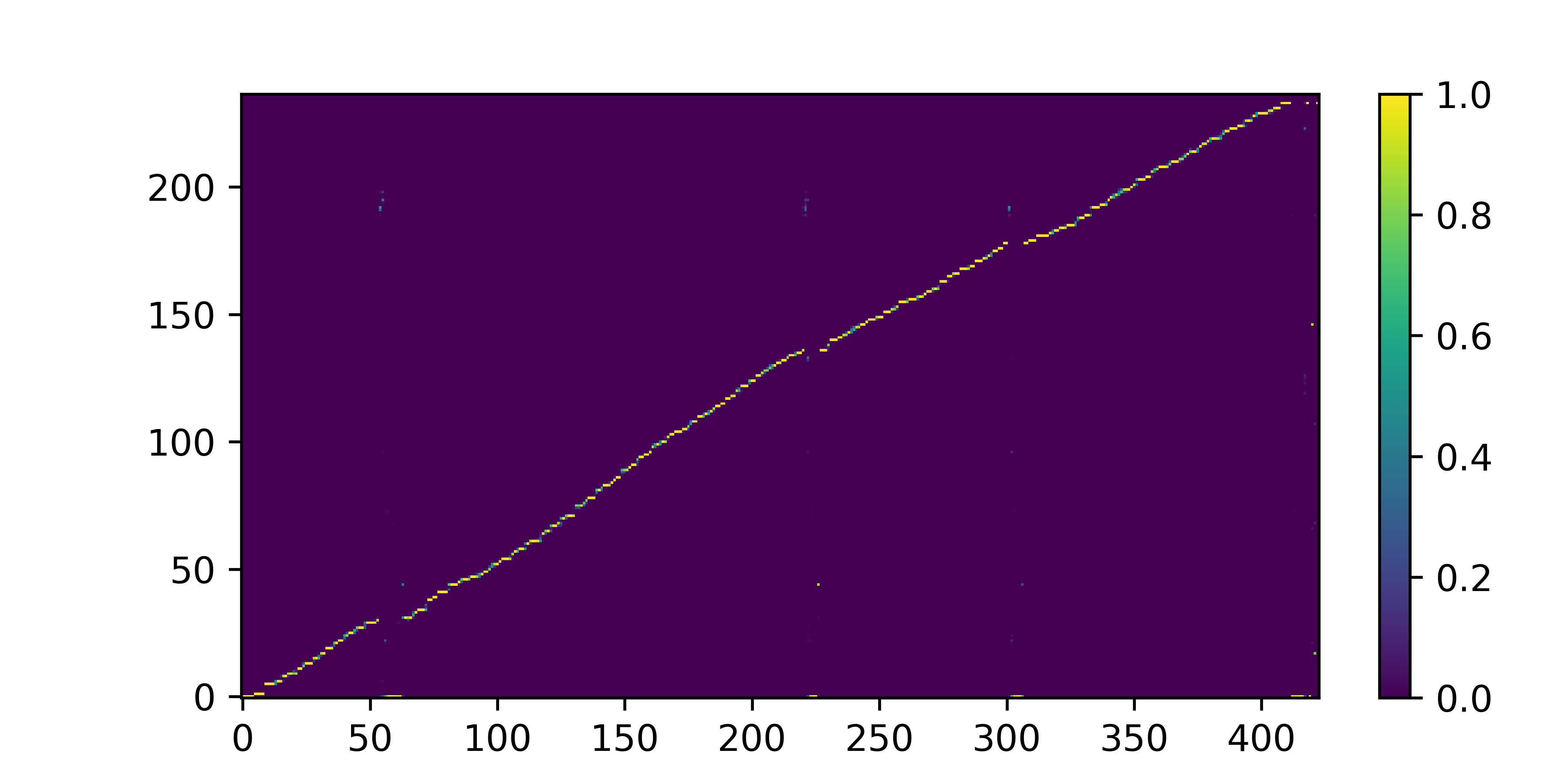}
 \caption{One of the learnt bidirectional attention weight plots, ${\bf W}_{12}$ in Eq.~\ref{eq:align-mat-after-softmat}. The horizontal axis represents speech embedding sequence ${\bf X}$, and the vertical axis is the text ones {\bf Y}. }
 \label{fig:biam-alignment}
\end{figure}

To see if the model has successfully learned the speech-to-text alignment with the help of the BiAM module, Figure~\ref{fig:biam-alignment} plots the alignment matrix after softmax operation, namely ${\bf W}_{12}$ in Equation~\ref{eq:align-mat-after-softmat}.
From Figure~\ref{fig:biam-alignment}, we can see the clear monotonic alignment pattern between the text and speech sequences, which again validates the effectiveness of the BiAM method.
In addition, the breakpoints in the alignment correspond to the ``Blank" label in Figure~\ref{fig:biam-alignment}. 

\section{Discussion \& Conclusion}\label{sec:discussion}
 The above experimental results show that the proposed bidirectional attention mechanism has clear advantages for speech-text forced-alignment learning, yielding improved ASR performance in a speech-text multi-modal training framework. 
However, the exploration is still far from perfect, and the limitations are at least as follows. 1) The effectiveness of the unpaired text pretraining is not fully demonstrated, particularly for full exploitation of the text data provided by Librispeech corpus is yet to be done.
2) Unpaired text pretraining method also needs a revisit in depth. So far, the pretraining is a mismatched one, yielding under-performed results. To realize a matched pretraining,
we need to figure out an approach to reconstruct the transform ${\bf W}_{12}$ in Eq.~\ref{eq:align-mat-after-softmat} for each unpaired utterance. Actually,
${\bf W}_{12}$ is not only ``diagonal" but also contains duration information for each grapheme. We are putting more effort on this in future.

% \section{Conclusion}~\label{sec:conclusion}
To conclude the work in this paper, we have proposed a speech-text-based multi-modal training framework for improving ASR performance via a bidirectional attention mechanism. We demonstrated its efficacy on Librispeech corpus with both 100- and 960-hour train data, respectively. With the paired speech-text-based multi-modal training, the proposed method has achieved up to 6.15\% and 5.66\% WER reductions on 4 test sets under the two scenarios. Besides,
on the 100-hour low-resource data, we also demonstrated the effectiveness of the proposed method for unpaired text data pretraining. 
Future work will be focused on efficient unpaired text data pretraining.

\vfill\pagebreak
\clearpage

% \bibliographystyle{IEEEbib}
\bibliographystyle{ieeetr}
\bibliography{refs}

\begin{thebibliography}{10}

\bibitem{chan2016listen}
W.~Chan, N.~Jaitly, {\em et~al.}, ``Listen, attend and spell: A neural network
  for large vocabulary conversational speech recognition,'' in {\em Proc. of
  ICASSP}, pp.~4960--4964, IEEE, 2016.

\bibitem{2017State_icassp}
C.-C. Chiu, T.~N. Sainath, {\em et~al.}, ``State-of-the-art speech recognition
  with sequence-to-sequence models,'' 2017.

\bibitem{zhang2020transformer}
Q.~Zhang, H.~Lu, H.~Sak, {\em et~al.}, ``Transformer transducer: A streamable
  speech recognition model with transformer encoders and rnn-t loss,'' in {\em
  Proc. of ICASSP}, pp.~7829--7833, IEEE, 2020.

\bibitem{INTERSPEECH-conformer-2020}
A.~Gulati, J.~Qin, {\em et~al.}, ``{Conformer: Convolution-augmented
  Transformer for Speech Recognition},'' in {\em Proc. Interspeech 2020},
  pp.~5036--5040, 2020.

\bibitem{zhang2022non}
C.-F. Zhang, Y.~Liu, {\em et~al.}, ``Non-autoregressive transformer with
  unified bidirectional decoder for automatic speech recognition,'' in {\em
  Proc. of ICASSP}, pp.~6527--6531, IEEE, 2022.

\bibitem{he2019streaming}
Y.~He, T.~N. Sainath, R.~Prabhavalkar, {\em et~al.}, ``Streaming end-to-end
  speech recognition for mobile devices,'' in {\em ICASSP}, IEEE, 2019.

\bibitem{sainath2020streaming}
T.~N. Sainath, Y.~He, B.~Li, {\em et~al.}, ``A streaming on-device end-to-end
  model surpassing server-side conventional model quality and latency,'' in
  {\em Proc. of ICASSP}, IEEE, 2020.

\bibitem{li2022language}
B.~Li, T.~N. Sainath, {\em et~al.}, ``A language agnostic multilingual
  streaming on-device asr system,'' {\em arXiv:2208.13916}, 2022.

\bibitem{li2022massively}
B.~Li, R.~Pang, Y.~Zhang, T.~N. Sainath, {\em et~al.}, ``Massively multilingual
  asr: A lifelong learning solution,'' in {\em Proc. of ICASSP}, IEEE, 2022.

\bibitem{ling2020deep}
S.~Ling, Y.~Liu, {\em et~al.}, ``Deep contextualized acoustic representations
  for semi-supervised speech recognition,'' in {\em Proc. of ICASSP},
  pp.~6429--6433, IEEE, 2020.

\bibitem{liu2020mockingjay}
A.~T. Liu, S.-w. Yang, {\em et~al.}, ``Mockingjay: Unsupervised speech
  representation learning with deep bidirectional transformer encoders,'' in
  {\em Proc. of ICASSP}, pp.~6419--6423, IEEE, 2020.

\bibitem{hsu2021hubert}
W.-N. Hsu, B.~Bolte, {\em et~al.}, ``Hubert: Self-supervised speech
  representation learning by masked prediction of hidden units,'' {\em IEEE/ACM
  Transactions on Audio, Speech, and Language Processing}, vol.~29,
  pp.~3451--3460, 2021.

\bibitem{zheng2021wav}
G.~Zheng, Y.~Xiao, {\em et~al.}, ``Wav-bert: Cooperative acoustic and
  linguistic representation learning for low-resource speech recognition,''
  {\em arXiv preprint arXiv:2109.09161}, 2021.

\bibitem{chung2019unsupervised}
Y.-A. Chung, W.-N. Hsu, {\em et~al.}, ``An unsupervised autoregressive model
  for speech representation learning,'' 2019.

\bibitem{baevski2019vq}
A.~Baevski, S.~Schneider, and M.~Auli, ``vq-wav2vec: Self-supervised learning
  of discrete speech representations,'' {\em arXiv preprint arXiv:1910.05453},
  2019.

\bibitem{baevski2020wav2vec}
A.~Baevski, Y.~Zhou, A.~Mohamed, {\em et~al.}, ``wav2vec 2.0: A framework for
  self-supervised learning of speech representations,'' {\em Advances in Neural
  Information Processing Systems}, vol.~33, pp.~12449--12460, 2020.

\bibitem{chen2022wavlm}
S.~Chen, C.~Wang, {\em et~al.}, ``Wavlm: Large-scale self-supervised
  pre-training for full stack speech processing,'' {\em IEEE Journal of
  Selected Topics in Signal Processing}, 2022.

\bibitem{baevski2022data2vec}
A.~Baevski, W.-N. Hsu, {\em et~al.}, ``Data2vec: A general framework for
  self-supervised learning in speech, vision and language,'' {\em arXiv
  preprint arXiv:2202.03555}, 2022.

\bibitem{chorowski2016towards_lm}
J.~Chorowski and N.~Jaitly, ``Towards better decoding and language model
  integration in sequence to sequence models,'' {\em arXiv preprint
  arXiv:1612.02695}, 2016.

\bibitem{sriram2017cold_lm}
A.~Sriram, H.~Jun, {\em et~al.}, ``Cold fusion: Training seq2seq models
  together with language models,'' {\em arXiv preprint arXiv:1708.06426}, 2017.

\bibitem{hori2019cycle}
T.~Hori, R.~Astudillo, T.~Hayashi, {\em et~al.}, ``Cycle-consistency training
  for end-to-end speech recognition,'' in {\em Pro. of ICASSP}, pp.~6271--6275,
  IEEE, 2019.

\bibitem{wang2020improving_tts}
G.~Wang, A.~Rosenberg, Z.~Chen, {\em et~al.}, ``Improving speech recognition
  using consistent predictions on synthesized speech,'' in {\em Proc. of
  ICASSP}, pp.~7029--7033, IEEE, 2020.

\bibitem{text-tts-2020-ibm}
Y.~Huang, H.-K. Kuo, S.~Thomas, {\em et~al.}, ``Leveraging unpaired text data
  for training end-to-end speech-to-intent systems,'' {\em arXiv:2010.04284},
  2020.

\bibitem{fb2021general}
Y.~Tang, J.~Pino, {\em et~al.}, ``A general multi-task learning framework to
  leverage text data for speech to text tasks,'' in {\em Proc. of ICASSP},
  IEEE, 2021.

\bibitem{2021Optimizing}
W.~Wang, S.~Ren, Y.~Qian, {\em et~al.}, ``Optimizing alignment of speech and
  language latent spaces for end-to-end speech recognition and understanding,''
  in {\em Proc. of ICASSP}, IEEE, 2021.

\bibitem{google2022MAESTRO}
Z.~Chen, Y.~Zhang, {\em et~al.}, ``Maestro: Matched speech text representations
  through modality matching,'' 2022.

\bibitem{li2022neufa}
J.~Li, Y.~Meng, Z.~Wu, {\em et~al.}, ``Neufa: Neural network based end-to-end
  forced alignment with bidirectional attention mechanism,'' in {\em Proc. of
  ICASSP}, pp.~8007--8011, IEEE, 2022.

\bibitem{jhu2018-mmda}
A.~Renduchintala, S.~Ding, {\em et~al.}, ``Multi-modal data augmentation for
  end-to-end asr,'' {\em arXiv:1803.10299}, 2018.

\bibitem{mit-2019-asru}
J.~Drexler and J.~Glass, ``Explicit alignment of text and speech encodings for
  attention-based end-to-end speech recognition,'' in {\em Proc. of ASRU},
  pp.~913--919, ASRU, 2019.

\bibitem{deng2022cif}
K.~Deng, S.~Cao, {\em et~al.}, ``Improving ctc-based speech recognition via
  knowledge transferring from pre-trained language models,'' in {\em Proc. of
  ICASSP}, pp.~8517--8521, IEEE, 2022.

\bibitem{speecht52022}
J.~Ao, R.~Wang, {\em et~al.}, ``Speecht5: Unified-modal encoder-decoder
  pre-training for spoken language processing,'' {\em arXiv:2110.07205}, 2021.

\bibitem{chen2022tts4pretrain}
Z.~Chen, Y.~Zhang, A.~Rosenberg, {\em et~al.}, ``Tts4pretrain 2.0: Advancing
  the use of text and speech in asr pretraining with consistency and
  contrastive losses,'' in {\em Proc. of ICASSP}, pp.~7677--7681, IEEE, 2022.

\bibitem{dong2020cif}
L.~Dong and B.~Xu, ``Cif: Continuous integrate-and-fire for end-to-end speech
  recognition,'' in {\em Proc. of ICASSP}, pp.~6079--6083, IEEE, 2020.

\bibitem{ICASSP-CTC-2018}
K.~Krishna, L.~Lu, K.~Gimpel, and K.~Livescu, ``A study of all-convolutional
  encoders for connectionist temporal classification,'' in {\em Proc. ICASSP
  2018}.

\bibitem{watanabe2018espnet}
S.~Watanabe, T.~Hori, S.~Karita, T.~Hayashi, J.~Nishitoba, Y.~Unno, N.~E.~Y.
  Soplin, J.~Heymann, M.~Wiesner, N.~Chen, {\em et~al.}, ``Espnet: End-to-end
  speech processing toolkit,'' {\em arXiv preprint arXiv:1804.00015}, 2018.

\end{thebibliography}

\end{document}